\documentclass[%
 aip,
 jcp,
 amsmath,amssymb,
 reprint,%
]{revtex4-2}

\usepackage[utf8]{inputenc}

\usepackage{siunitx}
\DeclareSIUnit\angstrom{\AA}
\usepackage[english]{babel}

\addto\captionsenglish{}

\usepackage{xifthen}
\usepackage{bm}
\usepackage{booktabs}
\usepackage{graphicx}
\usepackage{graphics}
\usepackage{xspace}
\usepackage{afterpage}

\let\svtextsl\textsl
\renewcommand\textsl[1]{\bgroup\fontfamily{ppl}\selectfont\svtextsl{#1}\egroup}
\let\svslshape\slshape
\renewcommand\slshape{\fontfamily{ppl}\selectfont\svslshape}
\usepackage[mathscr]{eucal}

\usepackage[dvipsnames]{xcolor}

\newcommand{\firstline}[2][]{%
  \ifthenelse{
    \equal{#1}{}}{\hspace{0em}&\hspace{-1em} {#2}\nonumber\\[0.5ex]
  }{
    \hspace{#1}&\hspace{-#1} {#2}\nonumber\\[0.5ex]
  }%
}

\usepackage{enumitem}
\newlist{resultlist}{enumerate}{1}
\setlist[resultlist, 1]{
    label=(\arabic{resultlisti}), 
    leftmargin=*,
    rightmargin=0pt,
    itemsep=0.75ex,
    parsep=0ex,
    topsep=0.75ex,
}
\usepackage[breaklinks]{hyperref}
\hypersetup{
    pdfauthor=author,
    colorlinks=true,
    linkcolor=cyan,
    citecolor=cyan,
    urlcolor=cyan,
}
\usepackage[capitalize]{cleveref}
  \crefformat{equation}{Eq.~(#2#1#3)}
  \crefrangeformat{equation}{Eqs.~(#3#1#4) to~(#5#2#6)}
  \Crefformat{equation}{Equation~(#2#1#3)}
  \Crefrangeformat{equation}{Equations~(#3#1#4) to~(#5#2#6)}

\newcounter{req}  

\crefname{req}{req}{req}
\crefname{req}{Req.}{Reqs.}
\creflabelformat{req}{(#1) #2 #3}

\crefname{inlineequation}{inlineequation}{inlineequation}
\crefname{inlineequation}{Req.}{Reqs.}
\creflabelformat{inlineequation}{(#1) #2 #3}

\DeclareMathAlphabet\mathbfcal{OMS}{cmsy}{b}{n}

\newcommand{\sm}{\textcolor{cyan}{supplementary material}\xspace}

\newcommand{\rv}{\mathrm{\textsl{v}}}

\newcommand{\mcd}{\mathcal{D}}

\newcommand{\bdot}[1]{\overset{\,_{\mbox{\large .}}}{#1}}

\newcommand{\CL}{\ensuremath{C_\parallel}}
\newcommand{\CT}{\ensuremath{C_{\!\!\:\perp}}}
\newcommand{\Fs}{\ensuremath{F_{\!s}}}
\newcommand{\CLdot}{\ensuremath{\dot{C}_\parallel}}
\newcommand{\CTdot}{\ensuremath{\dot{C}_{\!\!\:\perp}}}

\newcommand{\CLddot}{\ensuremath{\ddot{C}_\parallel}}
\newcommand{\CTddot}{\ensuremath{\ddot{C}_{\!\!\:\perp}}}

\newcommand{\DL}{\ensuremath{\mcd_\parallel}}
\newcommand{\DT}{\ensuremath{\mcd_\perp}}
\newcommand{\Ds}{\ensuremath{D_s}}
\newcommand{\vt}{\ensuremath{\rv_0}}
\newcommand{\vL}{\ensuremath{\rv_\parallel}}
\newcommand{\Np}{\ensuremath{\mathcal{N}_{\!p}{\!\:}}}

\newcommand{\dip}[1]{\ensuremath{d^\text{\!\;\tiny {#1}}}}
\newcommand{\taup}[1]{\ensuremath{\tau^\text{\!\:\tiny {#1}}}}

\let\baraccent=\= 
\newcommand{\emath}{\ensuremath}

\newcommand{\x}{\times}

\newcommand{\sdfrac}[2]{\mbox{\small$\displaystyle\frac{#1}{#2}$}}

\newcommand{\fk}[2][]{%
  \ifthenelse{\equal{#1}{}}{
    \emath{\hat{#2}_k}
  }{
    \emath{\hat{#1}_{#2}}
  }%
}

\renewcommand{\=}[1]{\stackrel{#1}{=}} 

\newcommand{\bigO}{\mathcal{O}}  

\DeclareRobustCommand{\v}[1]{\emath{\mathbf{#1}}}             
\DeclareRobustCommand{\gv}[1]{\emath{\mbox{\boldmath$ #1 $}}} 
\DeclareRobustCommand{\uv}[1]{\emath{\mathbf{\hat{#1}}}}      

\DeclareRobustCommand{\del}{\nabla}
\DeclareRobustCommand{\div}[1]{\del \cdot #1}

\DeclareRobustCommand{\intd}[1]{\int\!d{#1}\:\!}

\DeclareRobustCommand{\ii}[2]{\int_{#1}^{#2}}

\newcommand{\intzinf}[1][]{%
  \ifthenelse{\isempty{#1}}{
    \emath{\ii{0}{\infty}}
  }{
    \emath{\ii{0}{\infty}\!\!d{#1}\:\!}
  }%
}
\newcommand{\intinfinf}[1][]{%
  \ifthenelse{\isempty{#1}}{
    \emath{\ii{-\infty}{\infty}}
  }{
    \emath{\ii{-\infty}{\infty}\!\!d{#1}\:\!}
  }%
}

\newcommand{\stavg}[2][]{%
  \ifthenelse{\isempty{#1}}{
    \emath{ \left\langle{#2}\right\rangle }
  }{
    \emath{ #1{\langle}{#2}#1{\rangle} }
  }%
}
\newcommand{\abs}[1]{\left| #1 \right|} 

\DeclareRobustCommand{\kt}{\emath{k_BT}}
\DeclareRobustCommand{\msd}{\emath{\Delta}}

\newcommand{\pa}{\partial}

\newcommand{\fr}{\frac}

\newcommand{\be}{\begin{equation}}
\newcommand{\ee}{\end{equation}}

\newcommand{\bc}{\begin{center}}
\newcommand{\ec}{\end{center}}

\newcommand{\Sc}{\ensuremath{\mathrm{Sc}}}  

\newcommand{\citepunc}[2]{{\protect\NoHyper\citeauthor{#1}\protect\endNoHyper}#2\cite{#1}}

\makeatletter
\newcommand*{\citenst}[2][]{%
  \begingroup
  \let\NAT@mbox=\mbox
  \let\@cite\NAT@citenum
  \let\NAT@space\NAT@spacechar
  \let\NAT@super@kern\relax
  Ref.~\cite[#1]{#2}%
  \endgroup
}
\makeatother

\allowdisplaybreaks
\begin{document}


\title{Molecular hydrodynamic theory of the velocity autocorrelation function}

\author{S. L. Seyler}%
\email{slseyler@asu.edu}
\affiliation{
School of Molecular Sciences, Arizona State University, Tempe, AZ 85287
}%

\author{C. E. Seyler}%
\affiliation{
Laboratory of Plasma Studies, Cornell University, Ithaca, NY 14853
}%

\date{\today}

\begin{abstract}
    The velocity autocorrelation function (VACF) encapsulates extensive information about a fluid's molecular-structural and hydrodynamic properties. We address the following fundamental question: How well can a purely hydrodynamic description recover the molecular features of a fluid as exhibited by the VACF? To this end, we formulate a bona fide hydrodynamic theory of the tagged-particle VACF for simple fluids. Our approach is distinguished from previous efforts in two key ways: collective hydrodynamic modes are modeled by \emph{linear} hydrodynamic equations; the fluid's static kinetic energy spectrum is identified as a necessary initial condition for the momentum current correlation. Our formulation leads to a natural physical interpretation of the hydrodynamic VACF as a superposition of quasinormal hydrodynamic modes weighted commensurately with the static kinetic energy spectrum, which appears to be essential to bridging continuum hydrodynamical behavior and discrete-particle kinetics. Our methodology yields VACF calculations quantitatively on par with existing approaches for liquid noble gases and alkali metals; moreover, our hydrodynamic model for the self-intermediate scattering function extends the applicable domain to low densities where the Schmidt number is of order unity, enabling calculations for gases and supercritical fluids.

\end{abstract}


\maketitle 

\section*{Introduction}\label{sec:introduction}

The velocity autocorrelation function (VACF) is, perhaps, the quintessential time-correlation function, as it holds unique significance in condensed matter physics and chemistry.\cite{Boon1991-kx, McQuarrie2000-gi, Hansen2013-rh} In particular, the VACF is intrinsically linked to Brownian motion: the self-diffusion coefficient, $D_s$, of a large diffusing particle is the time integral of the VACF, a connection traceable to the famous Einstein relation, $\smash{D_s = \lim_{t\to\infty} \bigl\langle\abs{\v{x}(t)-\v{x}(0)}^2\bigr\rangle/6t}$.\cite{Einstein1905-lc} It was therefore unexpected when Alder and Wainwright,\cite{Alder1967-kk, Alder1970-kg} using molecular dynamics (MD) simulations, revealed that the VACF of a ``tagged'' fluid particle---one mechanically identical to all others---decays as $\smash{t^{-3/2}}$ at long times, not exponentially as predicted by Chapman-Enskog-Boltzmann theory.\cite{Enskog1917-sz, Enskog1922-sv, Brush1972-ro, Resibois1977-mh, Chapman1990-rl} Using simple physical arguments, they recognized that this protracted decay was due to delayed viscous momentum transport---a phenomenon known as \emph{hydrodynamic memory}.

\citet{Zwanzig1970-qy} promptly recognized that viscoelastic hydrodynamics could account for both hydrodynamic memory and molecular-scale granularity in the VACF. Although Zwanzig-Bixon theory remains questionable at short times, its success for liquid argon substantiates the hydrodynamic perspective.\cite{Schofield1975-sc, Boon1991-kx} Indeed, the fluid-particle VACF has broad utility as it probes molecular structure and dynamics across all timescales, yet efforts to obtain physically consistent analytic models spanning all scales and densities have been hampered by subtle challenges.\cite{Martin1968-hk, Forster1968-bp, Kim1971-ab, Pomeau1975-oa, Sjogren1979-fj, Gaskell1989-qk, Balucani1990-zu, Anento1999-bt} Meanwhile, MD simulation, along with methods based on the projection operator and memory function formalisms,\cite{Zwanzig1961-ao, Mori1965-za, Mori1965-rv} has been a workhorse for probing molecular scales\cite{Levesque1970-jo, Levesque1973-ue, Kushick1973-vs, Schofield1973-ca, *De_Schepper1984-hh, *McDonough2001-ny, *Dib2006-ot, *Sanghi2016-wh, Lesnicki2017-bm, Han2018-lc, *Mizuta2019-nl, Zhao2021-zu} while underpinning a burgeoning interest in multiscale modeling, including generalized Langevin equations (GLEs) and other coarse-graining (CG) techniques.\cite{Hijon2009-tu, *Espanol2009-cu, *Espanol2009-vf, *Morrone2011-jx, *Izvekov2013-tj, *Carof2014-ti, *Li2015-kj, Espanol2015-ks, Lesnicki2016-bc, Jung2016-mc, *Izvekov2017-du, *Jung2017-je, Jung2018-rw, *Rossi2018-lp, *Izvekov2019-ex, *Duque-Zumajo2020-yp, *Klippenstein2021-tw}

Nevertheless, complex fluid systems---nanocolloidal suspensions,\cite{Malevanets2000-dv, *Padding2004-sh, *Padding2006-im, *Dahirel2007-wi, *Wang2009-as, Espanol2015-ks, Jung2018-rw, *Bonaccorso2020-ia, *Moore2023-ff} active fluids and microswimmers,\cite{Koch2011-sv, *Wang2012-me, *Elgeti2015-sq, *Ghosh2015-rt, *Sharma2016-kh, *Szamel2019-hm, *Chakrabarti2023-wd, *Caprini2023-yb} solvated biomolecules,\cite{Garcia_De_La_Torre2000-kg, *Fernandes2002-ia, *Brangwynne2008-ao, *Kapral2016-or} ionic liquids and electrolyte solutions,\cite{Nee1970-jj, *Lantelme1979-hb, Gaskell1982-wv, *Malik2010-zh, *Gebbie2013-cf, *Wilkins2017-yf, *De_Souza2020-nt, *Ghorai2020-rn, *Sarhangi2020-na, *Samanta2021-nl, *Kournopoulos2022-sk}
and others\cite{Tang1996-jt, *Lisy2004-hc, *Howse2007-fy, *Bernabei2011-ax, *Huang2018-hv, *Gaspard2018-cj, *Szamel2022-vv}---exhibit a myriad of mesoscopic phenomena, including long-ranged hydrodynamic interactions and (active) Brownian motion, that computationally challenge particle-based simulations. And yet, it is not clear for even simple fluids\cite{Hansen2013-rh} what general rules delineate the applicable domain of continuum hydrodynamic models.\cite{Usabiaga2013-og, Han2018-lc} What ingredients are necessary to bridge continuum and discrete-particle behavior at molecular scales? And down to what scale can a hydrodynamic model yield a viable account of the VACF?

\begin{table*}[t!]
\centering
\caption{\label{tab:desiderata}Desiderata and constraints for physically admissible correlation functions.}
\begin{ruledtabular}
\renewcommand{\arraystretch}{1.15}
\begin{tabular}{@{}lll@{}}
Desideratum (req.)                                    & Expression                                                                    & Remarks                                                                       \\\hline
(1) memory equation\footnotemark[1]                   & $\bdot{C}  \!+ \!\int_0^t \! d\tau \:\! K(k,t\!-\!\tau)\:\!C(k,\tau) = 0$     & Memory kernel $\smash{K(k,t)}$ must be consistent with desiderata.                  \\
(2) normalizability\footnotemark[1]                   & $C(k,0) = 1$                                                                  & Static spectrum must be integrable: $\smash[b]{\intzinf[k]\!\!\: k^2q(ka)}$ exists.          \\
(3) time reversibility\footnotemark[1]                & $\smash{\bdot{C}(k,0) = 0}$                                                   & Process must be time-reversal symmetric.                                            \\
(4) exponential range\footnotemark[1]                 & $\smash{\CT(k,t) \sim e^{-t/\tau_\perp}}$, $\smash{\Fs(k,t) \sim e^{-t/\tau_s}}$    & Must decay exponentially for $\smash{k\vt\!>\!\tau_{\perp,s}^{-1}}$ at lower densities.\footnotemark[3] \\
(5) ballistic range\footnotemark[1]\footnotemark[2]   & $\smash{\psi(t) \sim \mathcal{N}\!\intzinf[k] k^2 q(ka) e^{-k^2\rv^2t^2/2}}$  & Short-time sum rule constrains $a$ using $\omega_E$ for liquids.\footnotemark[4]          \\
(6) diffusive range\footnotemark[1]\footnotemark[2]   & $\smash{\CT(k,t)\Fs(k,t) \sim e^{-k^2\left(\nu+D_s\right)\,t}}$               & Must preserve long-time asymptotic form of VACF.\footnotemark[5]                    \\
(7) long-time diffusivity\footnotemark[2]               & $\smash{D_s = \mathcal{N}\!\!\!\;\intzinf[t]\!\!\intzinf[k]\!\!\; k^2 q(ka)\CT(k,t)\Fs(k,t)}$  & Zero-frequency sum rule from Green-Kubo relation.\footnotemark[6]
\end{tabular}
\end{ruledtabular}
\raggedright
\footnotemark[1]{General requirements that constrain the functional forms of correlation functions.}\\
\footnotemark[2]{Specific requirements that constrain the values of correlation function parameters.}\\
\footnotemark[3]{Requires $\DT,\Ds \ne 0$ and $k$-independence of $\CT$ and $\Fs$ for large $k$. See \cref{sm:gas_parameters}.}\\
\footnotemark[4]{Sum rule constraint is valid when Einstein frequency, $\omega_E$, is well defined, e.g, for liquids. See \cref{sm:scale_a}.}\\
\footnotemark[5]{Implies the invariant forms $a^2\upsilon$ and $a^2\gamma_s$. See \crefrange{eq:asymptotic_decay}{eq:a_rescale} and \cref{eq:sm_gas_gamma_s,eq:sm_gas_a}.}\\
\footnotemark[6]{Liquids: $\CL \approx 0$, $\Fs \approx 1$ yields analytic solution for $\DT$. Gases: $\CL \approx 0$ yields constraint for $\gamma_s$, $\DT$. See \cref{sm:green_kubo_constraint,sm:gas_parameters}.}
\end{table*}

In this Communication, we formulate a theory of the VACF from a purely hydrodynamic standpoint that directly confronts these questions. Central to our framework is a general set of physical \emph{desiderata} (\cref{tab:desiderata}), which constrains a chosen hydrodynamic model so as to correctly reproduce both short-time molecular kinetics and collective dynamics across all timescales while minimizing ad hoc assumptions. Crucially, we find physical consistency requires that the equal-time velocity covariance be an integrable distribution over wavenumber, which we identify as an initial condition corresponding to the fluid's static kinetic energy spectrum. These considerations lead to a general theory representing a considerably different physical picture than the otherwise formally similar velocity-field approach.\cite{Ernst1970-gt, Ernst1971-tp, Schofield1975-sc, Gaskell1978-fy, Gaskell1978-gt, Gaskell1979-es}

To this end, we develop a 10-moment molecular-hydrodynamic model based on the well-established 13-moment equations,\cite{Grad1949-gf, Struchtrup2003-um, Ottinger2005-az, Ottinger2007-em, Jou2010-pr, Ottinger2010-dz, Struchtrup2013-hb, Torrilhon2015-lx} which originate as velocity moments of the linearized BGK-Boltzmann kinetic equation.\cite{Bhatnagar1954-gx, Zwanzig2001-bd, Klimontovich2012-as} The resulting moment-equations naturally yield wavenumber-dependent current correlations that reproduce, among other phenomena, molecular-scale viscoelasticity.\cite{Zwanzig1965-oc, Schofield1966-xq} By contrast, \emph{generalized hydrodynamics}\cite{Alder1984-bw, Boon1991-kx, Hansen2013-rh} begins with the Navier-Stokes equations, extending them to molecular scales by phenomenologically promoting transport coefficients to nonlocal quantities with frequency- and wavenumber-dependence. In practice, the velocity-field method has leveraged generalized hydrodynamics with notable success.\cite{Chung1969-sc, Alder1984-bw, Balucani1985-kr, Balucani1987-kg, Gaskell1982-wv, Balucani1990-hh, Anento1999-bt, Verdaguer2000-sa, Colangeli2009-bb, Garberoglio2018-dx}

The present theory offers important advantages, however. In particular, it implies a natural physical interpretation of the hydrodynamic VACF: a superposition of quasinormal hydrodynamic modes, each mode $k$ having a wavelength $2\pi/k$ and weight $\bigl\langle u_k^2\bigr\rangle$ proportional to the equilibrium probability density of fluid kinetic energy.
Moreover, unlike other approaches, the methodology enables realistic VACF calculations not only for simple liquids, but also \emph{gases} using the same underlying framework.
We apply the methodology to representative fluids---liquid rubidium and argon, and gaseous and supercritical argon---and find remarkable agreement with MD calculations.\cite{Schofield1973-ca, Balucani1984-if, Lesnicki2016-bc, Zhao2021-zu}

\section*{Theory of the Hydrodynamic VACF}\label{sec:framework}

Let $\v{u}(\v{x},t)$ be the Eulerian velocity field of a fluid derived as the velocity moment of an exact single-particle distribution function, $f(\v{x},\v{v},t)$, and $\v{x}_\alpha(t)$ be the position of constituent particle $\alpha$ of mass $m$. We then require the fluid velocity at $\v{x}_\alpha(t)$ to be equal to the particle velocity: $\smash{\v{u}[\v{x}_\alpha(t),t] = \v{v}_\alpha(t)}$, where $\smash{\v{v}_\alpha(t) \equiv \dot{\v{x}}_\alpha(t)}$. Expressing the Lagrangian velocity using Fourier components $\v{u}_\v{k}(t)$,
\begin{equation}
    \v{u}\bigl[\v{x}_\alpha(t),t\bigr] = \intd{\v{k}} \v{u}_\v{k}(t)e^{i\v{k}\cdot\v{x}_\alpha(t)}
\end{equation}
the velocity covariance $Q(t)\equiv \bigl\langle \v{v}_\alpha(0) \cdot \v{v}_\alpha(t) \bigr\rangle$ becomes
\begin{align}\label{eq:velocity_covariance}
  Q(t) &=
    \Bigl\langle
    \v{u}\bigl[\v{x}_\alpha(0),0\bigr] \cdot \v{u}\bigl[\v{x}_\alpha(t),t\bigr] \Bigr\rangle \nonumber\\
    &= \intd{\v{k}d\v{k}'} \Bigl\langle\v{u}_\v{k}(0)\cdot\v{u}_{\v{k}'}(t) \, e^{ i\v{k}\cdot\v{x}_\alpha(0) + i\v{k}'\cdot\v{x}_\alpha(t) } \Bigr\rangle \nonumber\\
    &\approx \intd{\v{k}} \Bigl\langle \v{u}^*_\v{k}(0)\cdot\v{u}^{}_\v{k}(t) \Bigr\rangle \Bigl\langle e^{ i\v{k}\cdot[\v{x}_\alpha(t)-\v{x}_\alpha(0)] }\Bigr\rangle
\end{align}
where in the last line we assume a spatially homogeneous system and decompose the ensemble average into a product of averages. Importantly, we assume \emph{a priori} uncorrelated particle displacements and fluid velocities whose underlying correlations will be reproduced by applying appropriate physical constraints (cf.~\cref{tab:desiderata}).

Assuming an isotropic fluid, we decompose the velocity covariance, $\smash{Q(k,t) \equiv \bigl\langle \v{u}^*_\v{k}(0)\cdot\v{u}^{}_\v{k}(t) \bigr\rangle = q(k)C(k,t)}$, where $\smash{q(k) \equiv Q(k,0)}$ and $\smash{C(k,t) = Q(k,t)/q(k)}$ is the (normalized) current correlation; $\smash{\Fs(k,t) \equiv \smash{\bigl\langle \exp\bigl\{i\v{k} \cdot \bigl[\v{x}_\alpha(t) - \v{x}_\alpha(0)\bigr]\bigr\} \bigr\rangle}}$ is the self-intermediate scattering function (SISF), the characteristic function for tagged-particle displacements. Altogether, this yields a general expression for the VACF
\begin{equation}\label{eq:VACF_integral}
    \psi(t) = \mathcal{N}\intzinf[k] k^2q(k)C(k,t)\Fs(k,t)
\end{equation}
where $\mathcal{N} \equiv \smash{\left[\intzinf[k] k^2 q(k)\right]^{-1}}$ is the normalization at $t=0$.

\subsection*{Equilibrium energy partitioning}

From \cref{eq:VACF_integral}, it is apparent that the customary statement of equipartition, $q(k) = \kt/m$, leads to a divergent integral. This is unsurprising, as a true white noise spectrum is unphysical. Thus, the fluid static kinetic energy spectrum, or \emph{spectral density} for short,
\begin{equation}\label{eq:static_spectrum}
    q(k) \equiv \Bigl\langle \abs{u_k(0)}^2 \Bigr\rangle
\end{equation}
must contain a microscopic parameter, $a$, that attenuates large-wavenumber contributions. In turbulence theory, a fundamental quantity analogous to $q(k)$ is the omnidirectional kinetic energy spectrum, $E(k)=2\pi\rho k^2q(k)$.\cite{Leslie1973-yx, Lvov1991-ac, Zhou2021-pp}

The inverse Fourier transform of $q(k)$ is formally identical to the ``form factor,'' $f(r)$, in the velocity-field method, wherein $a$ is \emph{a priori} treated as an effective molecular radius, which, in practice, is fixed near the mean intermolecular distance $\smash{(V/N)^{1/3} = n_0^{-1/3}}$. However, by only requiring consistency with the desiderata in \cref{tab:desiderata}---viz. reqs.~(2) and (5)---we deduce that $a$ more generally represents a kinetic correlation length, a scale above which a continuum description applies. Indeed, we find the \emph{shape} of $q(k)$, which must depend on the detailed intermolecular potential, strongly influences VACF oscillations, starkly contrasting with the form factor's largely peripheral role in ensuring normalizability. Note that \cref{eq:static_spectrum} implies $q(k)\ge 0$, whereas $\smash{\hat{f}(k)}$ is negative for a range of wavenumbers; conversely, $q(r) = \smash{\int_0^\infty dk\, k^2 q(k)\sin(kr)/kr}$ can oscillate radially about zero, thereby inducing oscillations in the VACF.

\subsection*{The hydrodynamic VACF}

For an isotropic fluid, the VACF decomposes into longitudinal and transverse components with respect to wavevector $\v{k}$ as $\psi(t) = \smash{\frac{1}{3}\psi_{\parallel}(t)+\frac{2}{3}\psi_{\perp}(t)}$; \cref{eq:VACF_integral} becomes
\begin{equation}\label{eq:hd_VACF}
  \psi(t) = \frac{\Np a^3}{3}\!\intzinf[k] k^2 q(ka)\Fs(k,t) \Bigl[\CL(k,t)+ 2\CT(k,t)\Bigr]
\end{equation}
where, after changing notation from $q(k)$ to $q(ka)$ and setting $\mathcal{N} = \Np a^3$, we have explicitly introduced $a$ and a parameter $p$ that controls the shape of $q(ka)$.

\Cref{eq:hd_VACF} and the desiderata in \cref{tab:desiderata} constitute a general formulation of the hydrodynamic VACF. To implement the framework, one specifies the static spectrum, $q(ka)$, and correlation functions, $C_{\parallel,\perp}(k,t)$ and $\Fs(k,t)$. In what follows, we develop hydrodynamic models satisfying reqs.~(1-7) in \cref{tab:desiderata} to obtain solutions for $C_{\parallel,\perp}(k,t)$ and $\Fs(k,t)$. We then introduce representative models for $q(ka)$ that yield simple-fluid VACFs in good agreement with MD calculations.

\section*{Hydrodynamic Model Formulation}\label{sec:formulation}

To describe \emph{collective} hydrodynamic modes, we develop a \emph{regularized} variant of the well-known 10-moment equations, a relaxation system of partial differential equations (PDEs) that we call \emph{R10}.\cite{Klimontovich2012-as} When appropriately constrained by the desiderata, R10 reproduces accurate VACFs for simple liquids, as well as realistic VACFs for intermediate-density fluids and gases. Note that R10 is not necessarily the optimal (or unique) hydrodynamic model consistent with the desiderata; in principle, our framework is compatible with other methodologies, including generalized hydrodynamics\cite{Akcasu1970-nx, Alder1984-bw, Boon1991-kx} and the GENERIC formalism.\cite{Struchtrup2003-um, Ottinger2005-az}

R10 may be obtained as a generalization of the BGK-Boltzmann equation using the moment method of hydrodynamics, where we consider two collisional relaxation rates, one each for the longitudinal and transverse components of the deviatoric stress tensor. We derive analytic expressions for longitudinal and transverse current correlations, which capture damped molecular-scale sound and elastic shear waves, respectively. In particular, the corresponding memory equations and kernels [req.~(1)] are \emph{not} assumed, as is commonly done, but rather implied by the R10 PDEs.

To describe the \emph{self}-motion of tagged particle $\alpha$, we propose a hydrodynamic model for the self-density, $n_s$, motivated by the R10 equations for collective hydrodynamics variables ($n$, $\v{j}$, and $\v{S}$). We then derive analytic expressions for the SISF via $\Fs(k,t) = \smash{\bigl\langle n^*_s(k,0)n_s(k,t)\bigr\rangle}$.

As there are no known exact solutions for $\Fs$, approximations have often leveraged the Gaussian assumption, which is exact in the small- and large-$k$ limits and, conveniently, directly relates to the mean-square displacement (MSD), e.g., $\Fs(k,t) = \smash{\exp\bigl[-\fr{1}{2}k^2\msd(t)\bigr]}$, where $\smash{\msd(t) \equiv \smash{\frac{1}{3}\bigl\langle|\v{x}_\alpha(t)-\v{x}_\alpha(0)|^2\bigr\rangle}}$ is the MSD of tagged particle $\alpha$. For instance, a cumulant expansion of $\Fs(k,t)$ with a Gaussian leading term directly reveals non-Gaussian effects, which are known to be relatively small for simple liquids.\cite{Chen1977-fr, Boon1991-kx} Unlike common approximations, our model for $\Fs$ satisfies reqs.~(1--6), which, along with our model for $\CT$, extends the description to low Schmidt number, $\smash{\Sc \equiv \nu/D_s \sim \bigO(1)}$, and reproduces the exponential decay range of gases---a feature not captured by simple diffusion (i.e., Fick's law) and other Gaussian models.

\subsection*{Regularized 10-moment model (R10)}

Consider the following linear transport equations
for an adiabatic equation of state:
\begin{gather}
  \pa_t\rho + \div\v{j} = 0 \label{eq:continuity}\\
  \pa_t\,\v{j} + \vL^2\del\rho + \div\v{S} = 0 \label{eq:momentum}\\
  \pa_t\v{S} + 2\vt^2 \dot{\gv{\varepsilon}} = \intd{\v{x}'}\Bigl[\mathbfcal{D}(\v{x}-\v{x}') \del'^2 -\gv{\Upsilon}(\v{x}-\v{x}')\Bigr]\cdot \v{S}(\v{x}',t)\label{eq:stress}
\end{gather}
where $\rho(\v{x},t)$ is the mass density, $\v{j}(\v{x},t)$ is the mass current density, $\v{S}(\v{x},t)$ is the deviatoric stress tensor, and $\dot{\gv{\varepsilon}} \equiv \smash{ \frac{1}{2}\bigl[\del\v{j} + \bigl(\nabla\v{j}\bigr)^\top - \frac{2}{3}\bigl(\div\v{j}\bigr)\!\:\v{1}\bigr] }$ is the rate-of-strain tensor; $\smash{\vt \equiv \sqrt{k_BT/m}}$ and $\vL$ are thermal and longitudinal phase velocities, respectively. We define the Fourier transform of the collision frequency tensor $\gv{\Upsilon}(\v{x})$ as
\begin{equation}
  \gv{\Upsilon}_\v{k} \equiv \uv{k}\uv{k}\!\:\gamma + \Bigl(\v{1} - \uv{k}\uv{k}\Bigr)\!\:\upsilon
\end{equation}
where $\smash{\uv{k}} \equiv \v{k}/k$, $\v{1}$ is the unit tensor, and $\smash{\uv{k}\uv{k}}$ and $\smash{\v{1} - \uv{k}\uv{k}}$ are longitudinal (normal) and transverse (shear) projection operators; $\gamma$ and $\upsilon$ are the respective collision frequencies. Similarly, for the stress diffusion tensor, $\mathbfcal{D}(\v{x})$,
\begin{equation}
  \mathbfcal{D}_\v{k} \equiv \uv{k}\uv{k}\!\:\DL + \Bigl(\v{1} - \uv{k}\uv{k}\Bigr)\!\:
  \DT
\end{equation}
with regularization diffusion coefficients $\DL$ and $\DT$.

Two points should be highlighted. First, the Fourier-transformed collision term, $\smash{-\gv{\Upsilon}_\v{k}\cdot\v{S}_\v{k}}$, is a relaxation approximation related to the BGK collision operator in the BGK-Boltzmann equation that gives rise to viscoelasticity: e.g., $\smash{\upsilon^{-1}}$ is the transverse component's Maxwell relaxation time. Second, the Fourier-transformed diffusive regularization term, $-\smash{k^2\mathbfcal{D}_\v{k} \cdot\v{S}_\v{k}}$, extends the description to higher Knudsen (Kn) number\cite{Struchtrup2003-um, Struchtrup2013-hb}---essential for molecular-scale fluid flows where $\smash{\text{Kn} \sim \bigO(1)}$ and conventional Navier-Stokes fails.\cite{Karniadakis2005-pv, Seyler2017-tm} Importantly, \emph{regularization implements moment closure} when constrained by the Green-Kubo relation for self-diffusion [req.~(7)], as well as eliminating spatial structure below physically meaningful scales [req.~(4)].

\subsection*{Relaxation limit of the stress components}

Working in Fourier space, the transverse projection of \cref{eq:momentum,eq:stress} yields
\begin{gather}
  \fr{d\v{j}_{\perp \v{k}}}{dt} = -i\v{k}\cdot\v{S}_{\perp\v{k}} \label{eq:momtk}\\
  \fr{d\v{S}_{\perp \v{k}}}{dt} + \bigl(\upsilon+k^2\DT\bigr)\v{S}_{\perp \v{k}} = -i\vt^2\!\:\v{k}\!\:\v{j}_{\perp \v{k}} \label{eq:stresstk}
\end{gather}
while the longitudinal projection of \crefrange{eq:continuity}{eq:stress} yields
\begin{gather}
  \fr{d\rho_\v{k}}{dt} = -i\v{k}\cdot\v{j}_{\parallel \v{k}} \label{eq:contk}\\
  \fr{d\v{j}_{\parallel \v{k}}}{dt} = -i \vL^2\,\v{k}\:\!\rho_\v{k} -i\v{k}\cdot\v{S}_{\parallel\v{k}}\label{eq:momlk}\\
  \fr{d\v{S}_{\parallel \v{k}}}{dt} + \bigl(\gamma + k^2\DL\bigr) \v{S}_{\parallel \v{k}} = -i\vt^2\left(\sdfrac{1}{3}\v{k}\!\:\v{j}_{\parallel\v{k}} + \v{j}_\v{k}\v{k} \right) \label{eq:stresslk}
\end{gather}
We then make the key assumption that the longitudinal stress relaxation rate, $\gamma + k^2\DL$, is faster than any other timescale, which circumvents solving a third-order differential equation for $\CL$ and proves to be a good approximation; \cref{eq:stresslk} relaxes to
\begin{equation}\label{eq:stresslk_relax}
  \v{k}\cdot\v{S}_{\parallel \v{k}} = -i\mu_k\,\v{kk}\cdot\v{j}_{\parallel \v{k}}
\end{equation}
where $\smash{\mu_k \equiv \fr{4}{3}\vt^2/\bigl(\gamma+k^2\DL\bigr)}$ is a $k$-dependent kinematic bulk viscosity. Here, $\gamma$ is determined in the small-$k$ limit from available data for the bulk viscosity, $\smash{\mu \equiv \vt^2/\gamma}$, and $\smash{\DL}$ is treated as a free parameter; the intuitive choice $\smash{\DL \to 4\mu/3}$ yields good results for present calculations.

It is instructive to note that the \emph{isotropic} relaxation limit (i.e., $\smash{\gamma=\upsilon}$, $\smash{\upsilon \gg \pa_t}$, and $\smash{\upsilon\gg k^2\DT}$) yields the linearized Navier-Stokes equations with \emph{steady-state} kinematic viscosity $\smash{\nu = \vt^2/\upsilon}$, where the stress equation relaxes to Newton's law of viscosity, $\v{S} = -2\nu\dot{\gv{\varepsilon}}$.

\subsection*{Modal current correlation functions}

Using \cref{eq:momtk,eq:stresstk} for the transverse component, and the relaxation approximation for the longitudinal component, \cref{eq:contk,eq:momlk,eq:stresslk_relax}, we derive ordinary differential equations (ODEs) in time for the modal correlation functions using $\v{j}_\v{k}(t) = \rho_0\v{u}_\v{k}(t)$, $\rho_0$ is the equilibrium mass density. For transverse current correlations, $\CT$, we combine \cref{eq:momtk,eq:stresstk} to obtain a second-order ODE for the transverse current
\begin{equation}\label{eq:jTk}
  \fr{d^2\v{j}_{\perp \v{k}}}{dt^2} + \bigl(\upsilon+k^2\DT\bigr)\fr{d\v{j}_{\perp \v{k}}}{dt}+k^2\vt^2\,\v{j}_{\perp \v{k}}=0
\end{equation}
which leads to
\begin{equation}\label{eq:CT_ode}
  \CTddot(k,t) + \upsilon_k\CTdot(k,t) + k^2\vt^2\,\CT(k,t)=0
\end{equation}
where $\upsilon_k \equiv \upsilon + k^2\DT$. \Cref{eq:CT_ode} has a memory equation form, req.~(1), with kernel $K_\perp(k,t) = k^2\vt^2 \exp(-\upsilon_kt)$. Similarly, substitution of the relaxation limit, \cref{eq:stresslk_relax}, into \cref{eq:momlk} eventually yields a second-order ODE for the longitudinal current correlations
\begin{equation}\label{eq:CL_ode}
  \CLddot(k,t) + \gamma_k\CLdot(k,t) + k^2 \vL^2\, \CL(k,t)=0
\end{equation}
where $\gamma_k\equiv k^2\mu_k$. With the initial conditions $C_{\parallel,\perp}(k,0) = 1$ and $\dot{C}_{\parallel,\perp}(k,0) = 0$ [reqs.~(2--3)], solutions to \cref{eq:CT_ode,eq:CL_ode} are readily found:
\begin{gather}
  \CT(k,t) = \sdfrac{1}{2}e^{-\fr{1}{2}\upsilon_k t}\left[\biggl(1 - \sdfrac{\upsilon_k}{\alpha_k}\biggr)\!\:e^{-\fr{1}{2}\alpha_kt} + \biggl(1 + \sdfrac{\upsilon_k}{\alpha_k}\biggr)\!\:e^{\fr{1}{2}\alpha_kt}\right] \label{eq:CTk}\\
  \CL(k,t) = \sdfrac{1}{2}e^{-\fr{1}{2}\gamma_k t}\left[\biggl(1 - \sdfrac{\gamma_k}{\beta_k}\biggr)\!\:e^{-\fr{1}{2}\beta_kt} + \biggl(1 + \sdfrac{\gamma_k}{\beta_k}\biggr)\!\:e^{\fr{1}{2}\beta_kt}\right] \label{eq:CLk}
\end{gather}
where $\alpha_k \equiv \sqrt{\smash[b]{\upsilon_k^2} - \smash[b]{4k^2\vt^2}}$ and $\beta_k\equiv \sqrt{\smash[b]{\gamma_k^2}-\smash[b]{4k^2\vL^2}}$. \Cref{eq:CTk,eq:CLk} have the flexibility to satisfy all requirements in \cref{tab:desiderata}; the corresponding dynamic spectral densities are provided in \cref{sm:dynamic_spectral_density}.

\subsection*{Self-intermediate scattering function}

The regularized relaxation models represented by \cref{eq:momtk,eq:stresstk} and \crefrange{eq:contk}{eq:stresslk} for the collective hydrodynamic variables suggest the following model
\begin{gather}
    \fr{dn_s}{dt} = -i\v{k}\cdot\v{j}_s \label{eq:self_density}\\
    \fr{d\v{j}_s}{dt} + \bigl(\gamma_s+k^2D_s\bigr)\!\;\v{j}_s = -i\v{k}\vt^2 n_s \label{eq:self_current}
\end{gather}
where $n_s(k,t)$ is the (self-)density of a \emph{non-interacting} collection of tagged particles (i.e., test particles\cite{Chen1977-fr}), $\v{j}_s(k,t)$ the self-current, $\gamma_s$ the Brownian collision frequency corresponding to (Stokes) friction, and $D_s$ the self-diffusion coefficient; here, we take $\smash{D_s \to \vt^2/\gamma_s}$, which enforces req.~(4) and ensures the positivity of $\Fs$ for all $k$ and $t$.

\Cref{eq:self_density,eq:self_current} are a natural generalization of Fick's Law of self-diffusion,\cite{Ernst1970-gt,Ernst1971-tp} which is recovered in the full relaxation limit ($\smash{\gamma_s\to\infty}$, $\smash{k \to 0}$) where \cref{eq:self_density,eq:self_current} reduce to a diffusion equation for $n_s(k,t)$. The full ODE corresponding to \cref{eq:self_density,eq:self_current} is
\begin{equation}\label{eq:SISF}
    \ddot \Fs(k,t) + \bigl(\gamma_s+k^2D_s\bigr)\dot \Fs(k,t)+k^2\vt^2\Fs(k,t) = 0
\end{equation}
the roots of which factor nicely to give
\begin{equation}\label{eq:SISF_equation}
    \Fs(k,t) = \fr{\gamma_s e^{-k^2D_st}-k^2D_se^{-\gamma_s t}}{\gamma_s-k^2D_s}
\end{equation}
Note that density fluctuations decay exponentially for large $k$. To see that this model is reasonable, consider \cref{eq:SISF} in memory equation form
\begin{equation}
    \dot \Fs(k,t) + k^2\vt^2\int_0^t\!\!dt'e^{-\gamma_{sk}(t-t')}\Fs(k,t') = 0
\end{equation}
where $\gamma_{sk} = \gamma_s+k^2D_s$, $\gamma_s>0$. The Markovian solution, obtained by freezing $\Fs(k,t')$ at the upper limit $t$, is
\begin{equation}\label{eq:Markovian_SISF}
    \Fs(k,t) = \exp\left[-k^2\biggl(\sdfrac{\vt}{\gamma_{sk}}\biggr)^2\Big(\gamma_{sk}t+e^{-\gamma_{sk}t}-1\Big)\right]
\end{equation}
which, when $\smash{D_s = 0}$, yields the formal result for conventional Langevin dynamics.\cite{Boon1991-kx} Also note that when $\smash{D_s \to 0}$ in \cref{eq:SISF_equation}, $\smash{F_s(k,t) \to 1}$ and \cref{eq:Markovian_SISF} would not satisfy req.~(4). It is unsurprising that $\smash{D_s > 0}$ is necessary for a physically meaningful SISF.


\section*{Implementation of the Framework}\label{sec:implementation}

The VACF is calculated by evaluating \cref{eq:hd_VACF} using the transverse and longitudinal current correlations, \cref{eq:CTk} and \cref{eq:CLk}, and SISF, \cref{eq:SISF_equation}, along with the static spectrum, $q(ka)$, discussed subsequently; importantly, all VACF calculations use these equations (and parameters). However, determining gas parameters (\cref{sm:gas_parameters}) is considerably more complicated since the short-time sum rule [req.~(5)] only applies to dense fluids, $\Fs \approx 1$ cannot be assumed for req.~(7), and empirical data for transport parameters is limited. A suitable kinetic theory, such as Enskog theory (used here) or modifications thereof,\cite{Van_Beijeren1973-ml, Van_Beijeren1973-rb, Karkheck1981-cf} can be used with our framework to derive sum rules connecting molecular parameters to macroscopic quantities,\cite{Resibois1977-mh, Henderson1992-zk} as well as estimating transport coefficients in lieu of empirical inputs.

\subsection*{Static spectrum}\label{sec:static_spectrum}

Pending a first-principles derivation of $q(ka)$, we consider representative two-parameter models: a symmetric generalized Gaussian
\be\label{eq:gauss_spectrum}
    q_G(ka) = \mathcal{N}^G \vt^2 e^{-\frac{1}{2}|ka|^p}
\ee
with $\smash{p \geq 1}$, and generalized Lorentzian
\be\label{eq:lorentz_spectrum}
    q_L(ka) = \mathcal{N}^L \frac{\smash{\vt^2}}{1+|ka|^p}
\ee
with $\smash{p \geq 6}$, where the normalizations depend on shape parameter $p$ via $\smash{\mathcal{N}^G \equiv \mathcal{N}^G_{\!p} a^3}$ and $\smash{\mathcal{N}^L \equiv \mathcal{N}^L_{\!p} a^3}$. \Cref{eq:gauss_spectrum,eq:lorentz_spectrum} approach the equipartition spectrum for small $a$; softer intermolecular potentials correspond to sharper spectral cutoffs (larger $p$), which amplify oscillations in the VACF [cf.~\cref{fig:liquid_rubidium_vacf}].\cite{Schiff1969-lm, Geszti1976-mx, Canales1997-tk, Canales1999-xs} Note that normalization [req.~(2)] yields $\smash{\mathcal{N}^G_{\!p} = p/[8^{1/p}\Gamma(3/p)}]$, whereas solutions for $\mathcal{N}^L_{\!p}$ are unavailable for general $p$, so \cref{eq:gauss_spectrum} is preferred for analytic manipulation.

\subsection*{Parameter determination for liquids}\label{sec:param_liquid}

\begin{table}[t]
\centering
\caption{\label{tab:parameters}Key parameters. Left two columns: relations and constraints for molecular parameters. Right three columns: input data for liquid rubidium/argon calculations in \cref{fig:liquid_rubidium_vacf,fig:liquid_argon_vacf}.}
\begin{ruledtabular}\renewcommand{\arraystretch}{1.10}
\begin{tabular}{@{}crcrr@{}}
 Param.                           & Expr.             & Input              & \multicolumn{2}{c}{Value}                                                                                                                                    \\
\cmidrule(lr){4-5}
                                  &                   &             & \multicolumn{1}{c}{rubidium}                                                       & \multicolumn{1}{c}{argon}                                                        \\\hline
$\nu$                             &  $\eta/\rho_0$    & $T$         & \SI{319.}{\kelvin}\,\cite{Rahman1974-yq}                                          & \SI{87.}{\kelvin}\,\cite{Gaskell1978-gt}                                        \\
$\mu$                             &  $\eta_b/\rho_0$  & $\rho_0$    & \SI[per-mode=symbol]{1500.}{\kilogram\per\cubic\meter}\,\cite{Rahman1974-yq}      & \SI[per-mode=symbol]{1430.}{\kilogram\per\cubic\meter}\,\cite{Gaskell1978-gt}   \\
$\upsilon_0$                      &  $\vt^2/\nu$      & $\eta$      & \SI{644.}{\micro\pascal\second}\,\cite{Andrade1952-lw}                            & \SI{290.}{\micro\pascal\second}\,\cite{Lemmon2004-oe}                           \\
$\gamma_0$                        &  $\vt^2/\mu$      & $\eta_b$    & \SI{2020}{\micro\pascal\second}\,\cite{Zaheri2003-vk}                             & \SI{90.}{\micro\pascal\second}\,\cite{Chatwell2020-wc}                          \\
$\gamma_{s0}$                     &  $\vt^2/D_s$      & $D_s$       & \SI[per-mode=symbol]{2.85e-9}{\meter\squared\per\second}\,\cite{Van_Loef1974-yb}  & \SI[per-mode=symbol]{1.75e-9}{\meter\squared\per\second}\,\cite{Fincham1983-ua} \\
$a$                               &  req.~(5)         & $\omega_E$  & \SI{6.1}{\per\pico\second}\,\cite{Gaskell1978-fy}                                 & \SI{7.7}{\per\pico\second}\,\cite{Schofield1973-ca}                             \\
$\upsilon$, $\gamma$, $\gamma_s$  &  req.~(6)         & $\vL^2$     & $1.8\vt^2$\,\footnotemark[1]                                                       & $2.4\vt^2$\,\footnotemark[1]                                                     \\
$\DT$                             &  req.~(7)         & $p$         & 14\,\footnotemark[1]                                                               & 2\,\footnotemark[1]                                                              \\
$\DL$                             &  $4\mu/3$         & $q(ka)$     & $q_L(ka)$\,\footnotemark[2]                                                        & $q_G(ka)$\,\footnotemark[2]                                                      \\
\end{tabular}
\end{ruledtabular}
\raggedright
\footnotemark[1]{Free model parameter}\\
\footnotemark[2]{See \cref{eq:lorentz_spectrum,eq:gauss_spectrum}.}
\end{table}
%

The short- [req.~(5)] and long-time [req.~(6)] behavior of the VACF uniquely determines $a$ and relaxation frequencies $\upsilon$ and $\gamma_s$. To see this, consider the long-time behavior by evaluating \cref{eq:hd_VACF} with the asymptotic forms $\CT(k,t) \sim \smash{\exp\bigl[-k^2(\vt^2/\upsilon) t\bigr]}$ and $\Fs(k,t) \sim \smash{\exp\bigl[-k^2(\vt^2/\gamma_s) t\bigr]}$
\begin{align}
  \psi(t)
    &\sim \frac{2}{3}\Np a^3\!\intzinf[k] k^2 q(ka) e^{-k^2(\vt^2/\gamma_s) t} e^{-k^2(\vt^2/\upsilon) t} \\
    &\sim \frac{\sqrt{\pi}\Np}{6}\left[\frac{\vt^2}{a^2}\biggl(\frac{1}{\upsilon}+\frac{1}{\gamma_s}\biggr)\,t\right]^{-3/2}\label{eq:asymptotic_decay2}
\end{align}
along with the theoretical prediction from mode-coupling\cite{Ernst1970-gt, Ernst1971-si, Ernst1971-tp} and kinetic theory\cite{Dorfman1970-jw, Dorfman1975-va}
\begin{equation}\label{eq:asymptotic_decay}
    \psi(t) \sim  \frac{2}{3n_0}\left[\frac{1}{4\pi(\nu +D_s)\,t}\right]^{3/2}
\end{equation}
Naively, one can match \cref{eq:asymptotic_decay} by assuming $\smash{\upsilon \to \vt^2/\nu}$, $\smash{\gamma_s \to \vt^2/\Ds}$, and $\smash{\sqrt{\pi}\Np a^3/6 \to (2/3n_0) (4\pi)^{-3/2}}$. This satisfies req.~(6) and implies $\smash{a \to a_0 \equiv\bigl( 2\pi^2\Np n_0\bigr)^{-1/3}}$. But the short-time sum rule [req.~(5)] generally yields $a \leq a_0$ (cf.~\cref{sm:scale_a}), implying that the molecular-scale parameters $\smash{\upsilon \neq \vt^2/\nu}$ and $\smash{\gamma_s \neq \vt^2/\Ds}$. Assigning consistent values for $a$, $\upsilon$, and $\gamma_s$ therefore requires a more careful procedure to concomitantly satisfy reqs.~(5) and (6). Importantly, treating the products $a^2\upsilon$ and $a^2\gamma_s$ in \cref{eq:asymptotic_decay2} as invariant quantities preserves the asymptotic form.

These observations suggest the following procedure. First, define the following ``base'' values: $\smash{\upsilon_0 \equiv \vt^2/\nu}$ and $\smash{\gamma_{s0} \equiv \vt^2/\Ds}$ (cf.~\cref{tab:parameters}). Second, determine $a$ through the short-time sum rule. E.g., for a generalized Gaussian spectrum, one obtains (cf.~\cref{sm:scale_a})
\begin{equation}\label{eq:a_rescale}
    a^2 = \frac{4^{1/p}\Gamma(5/p)}{\Gamma(3/p)} \x \frac{1}{3\omega_E^2}\Bigl(\vL^2+2\vt^2\Bigr)
\end{equation}
Finally, take $\smash{\upsilon \to (a_0/a)^2 \upsilon_0}$ and $\smash{\gamma_s \to (a_0/a)^2 \gamma_{s0}}$, which preserves \cref{eq:asymptotic_decay}. Consistency suggests the longitudinal collision frequency be rescaled accordingly: $\smash{\gamma \to (a_0/a)^2 \gamma_0}$, where $\smash{\gamma_0 \equiv \vt^2/\mu}$.

\subsection*{Parameter determination at lower densities}\label{sec:param_low_density}

Given the paucity of numerical and experimental data for dilute monatomic fluids, we determine inputs using Enskog theory supplemented by heat capacity data from the National Institute of Standards and Technology (NIST) Chemistry WebBook.\cite{Lemmon2004-oe} For brevity, procedural details are provided in \cref{sm:gas_parameters}.

\section*{Numerical results}\label{sec:results}

\begin{figure}[t]
  \sisetup{per-mode=symbol}
  \centering
  \includegraphics[width=\columnwidth]{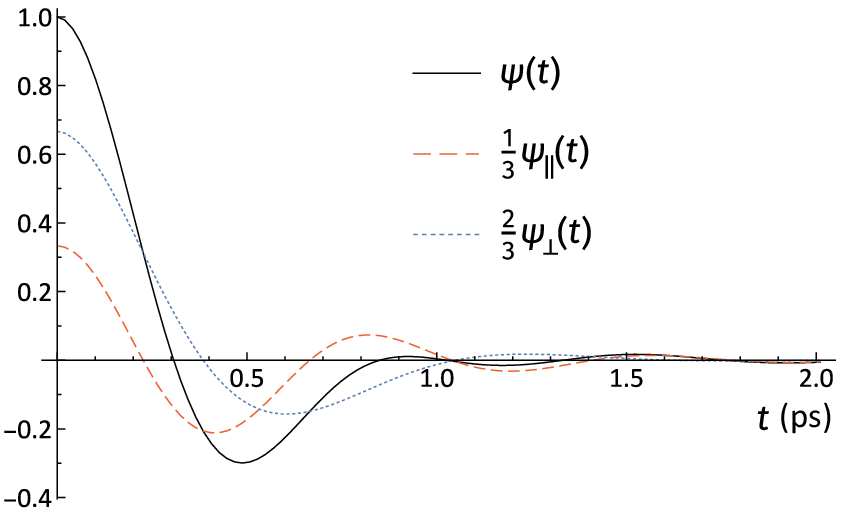}
  \caption{Liquid rubidium VACF ($\smash{T\approx\SI{319}{\kelvin}}$, $\rho_0 \approx \SI{1.51e3}{\kilo\gram\per\cubic\meter}$) with longitudinal (orange) and transverse (blue) components, using a generalized Lorentzian static spectrum ($p=14$).}
  \label{fig:liquid_rubidium_vacf}
\end{figure}
VACF calculations for liquid rubidium and argon use values listed in \cref{tab:parameters}. Calculations for argon-like gaseous and supercritical fluids use expressions for the Enskog viscosity and diffusion coefficients combined with reqs.~(4) and (7) to derive reasonable values for the key parameters $\smash{\DT}$, $\upsilon$, $\smash{\gamma_s}$, and $a$ (cf. \cref{sm:gas_parameters}). All calculations in presented in the article and its appendices can be performed with the Mathematica notebook provided in the \sm.

\Cref{fig:liquid_rubidium_vacf} shows the VACF for liquid rubidium using a generalized Lorentzian spectrum with $\smash{p=14}$, which should be compared to the velocity-field results (\citenst{Gaskell1978-fy}, Fig.~1; \citenst{Balucani1984-if}, Fig.~2). The oscillatory VACF behavior of liquid alkali metals observed in MD simulations\cite{Gaskell1978-fy, Gaskell1978-gt, Balucani1984-if, Balucani1992-qb} requires a relatively sharp cutoff (large $p$), which corresponds to the relatively soft repulsive core of the intermolecular potential as compared to liquid argon.\cite{Schiff1969-lm, Geszti1976-mx, Anento1999-bt}
\begin{figure}[t]
  \sisetup{per-mode=symbol}
  \centering
  \includegraphics[width=\columnwidth]{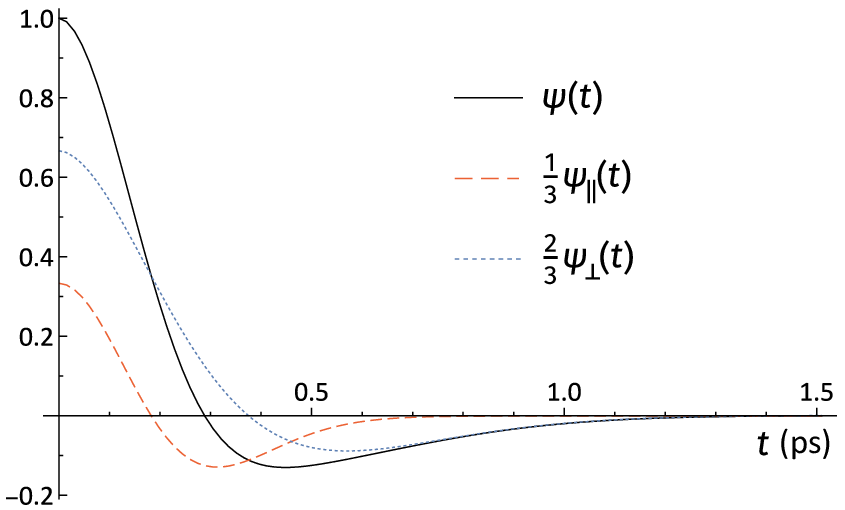}
  \caption{Liquid argon VACF near the triple point ($\smash{T\approx\SI{87}{\kelvin}}$, $\smash{\rho_0 \approx \SI{1.43e3}{\kilo\gram\per\cubic\meter}}$) with longitudinal (orange) and transverse (blue) components, using a Gaussian static spectrum ($p=2$).}
  \label{fig:liquid_argon_vacf}
\end{figure}
\Cref{fig:liquid_argon_vacf} shows the VACF for liquid argon near the triple point using a Gaussian ($\smash{p=2}$) spectrum (cf. velocity-field result: \citenst{Gaskell1978-gt}, Fig.~2). We remark that the plateau often seen in liquid argon(-like) VACFs from MD\cite{Rahman1964-ui, Levesque1970-jo, Schofield1973-ca, Levesque1973-ue, Kushick1973-vs, Fincham1983-ua, Meier2004-yo, Kim2015-wg} can be reasonably reproduced in our formulation using, e.g., a generalized Lorentzian spectrum with $\smash{p \approx 7.5}$, $\smash{\rv^2_\parallel \approx 3.0\vt^2}$, and adjusting $D_s$ upward by ${\sim}10\%$ so as to increase oscillations primarily in the \emph{longitudinal} (but not transverse) component.

\Cref{fig:gaseous_argon} shows VACFs for argon-like gaseous and supercritical fluids. The exponential and diffusive ($t^{-3/2}$) decay ranges are a direct consequence of the explicit handoff in \cref{eq:SISF_equation} and \cref{eq:sm_factored_CT}. These prominent features were observed in MD calculations for hard-spheres \cite{Zhao2021-zu} and a supercritical Lennard-Jones (LJ) fluid,\cite{Lesnicki2016-bc} as well as immersed-particle fluctuating hydrodynamics simulations at low Sc\cite{Usabiaga2013-dg, Usabiaga2013-og, Balboa_Usabiaga2014-rs} and analytic calculations for Basset-Boussinesq-Oseen (BBO) dynamics\cite{Boussinesq1885-tj, Basset1887-hh, Oseen1927-qd, Zwanzig1970-qy, Maxey1983-gz} with general slip boundary conditions.\cite{Gaspard2019-bl} In particular, the dips at early times (\cref{fig:gaseous_argon}, $\smash{t \lesssim \SI{10}{\pico\second}}$) come from the longitudinal current, which appears to capture the effect of strongly damped sound waves.
Similar features are clearly evident in MD calculations: see \citenst{Zhao2021-zu}, Fig.~S1 and also \cref{sm:sc_argon} for an indirect comparison with \citenst{Lesnicki2016-bc}, Fig.~1, which shows excellent quantitative agreement. Analytic calculations for ``BBO particles''\cite{Seyler2019-dg} also exhibit qualitative similarities (\citenst{Gaspard2019-bl}, Figs.~2 and 3).

\afterpage{
\begin{figure}[t]
  \sisetup{per-mode=symbol}
  \centering
  \includegraphics[width=\columnwidth]{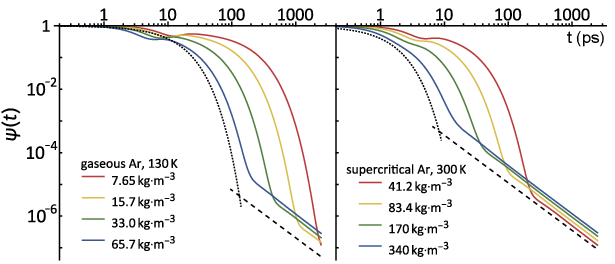}
  \caption{Hard-sphere VACFs emulating gaseous (left) and supercritical (right) argon ranging from low (red) to high (blue) densities. Pure exponential (dotted) and $t^{-3/2}$ (dashed) decay are shown as guides to the eye. Note that at \SI{300}{\kelvin}, \SI{41.2}{\kilogram\per\cubic\meter} (right, red) argon is \emph{gaseous} (and not supercritical).}
  \label{fig:gaseous_argon}
\end{figure}
}

\section*{Discussion}\label{sec:conclusion}

The present theory reflects an eclectic synthesis of ideas dispersed throughout the literature along with several new concepts. We have organized the theoretical formulation and framework in this Communication so as to highlight important results.
In summary, we:
\begin{enumerate}[
  label=(\arabic*),
  leftmargin=*,
  rightmargin=0pt,
  itemsep=0.35ex,
  parsep=0ex,
  topsep=0.35ex
]
    \item Presented a new derivation and interpretation of the hydrodynamic VACF formulation, \crefrange{eq:velocity_covariance}{eq:hd_VACF}.
    \item Established a core set of physical desiderata whose constraints are sufficient to recover realistic VACFs.
    \item Identified $q(ka)$ as the initial condition of the velocity covariance function that characterizes molecular-scale kinetic fluctuations and reproduces the VACF over \emph{all} timescales by judiciously superimposing each hydrodynamic mode.
    \item Proposed linear PDEs (R10) to model hydrodynamic collective modes, where regularization and the zero-frequency sum rule effect moment closure; yields analytic solutions for $\CT$ and $\CL$ with sufficiently rich structure to resolve subtle details in the VACF.
    \item Proposed a new hydrodynamic model for the self-density, leading to a viable analytic form of the self-intermediate scattering function, $\Fs(k,t)$, for all densities; captures exponential decay at low densities.
    \item Described the (re-)scaling of hydrodynamic model parameters that recover the short-time VACF behavior while preserving the correct long-time $t^{-3/2}$ decay.
\end{enumerate}
It is worth mentioning that one can derive telegrapher's equations from R10 (for $\smash{\DT}$ or $\smash{\Ds=0}$). \citet{Trachenko2017-mv} derived a telegrapher's equation as a continuum liquid dynamics model. We remark that telegrapher's equations describe \emph{persistent} random walks by accounting for directional correlations in Brownian motion.\cite{Furth1917-xp, Taylor1922-ro, Goldstein1951-dx} As pointed out by \citepunc{Khrapak2021-nz}{,} Zwanzig's speculative model of molecular self-motion in a liquid\cite{Zwanzig1983-kt}---where collective rearrangements correspond to configurational transitions between metastable equilibria---is consistent with the persistent random walk picture. From our hydrodynamic standpoint, this persistence originates from the finite relaxation time of molecular-scale stresses.

Regularization, however, is essential---especially at low densities. Importantly, \cref{eq:SISF_equation} shows that exponential decay occurs when $\smash{k^2D_s>\gamma_s}$, which implies $\smash{k\lambda^{}_E>1}$, where $\smash{\lambda^{}_E\sim\vt/\gamma_s}$ is the Enskog mean free path. More precisely, when $\smash{a< k^{-1} < \lambda^{}_E}$, the expected exponential decay range of a gas emerges from the contributions of large-$k$ modes. Also, note that the slowest relaxation rate of $\CT(k,t)$ is given by the smaller of the two exponents in \cref{eq:CTk}, which, when expanded for large $k$, gives $\smash{-\vt^2/\DT}$ when $\smash{\DT>0}$ (cf. \cref{sm:gas_parameters}). Thus, the regularization coefficients $\DT$ and $D_s$, the latter of which arises from our hydrodynamic model for the self-density, are necessary for obtaining the expected exponential decay of the dilute gas VACF.

The exponential decay of the SISF (and VACF) at low densities also hints at a deeper connection to VACFs for large Brownian particles, which also exhibit exponential decay.\cite{Balucani1994-mp} It is worth exploring the connection between our equations of motion for the self-density and GLE models of (single-)particle dynamics---particularly the fluctuating BBO equation, which describes hydrodynamic Brownian motion.\cite{Chow1972-mo, Nelkin1972-cp, Seyler2020-kj} Recent developments in memory kernel reconstruction methods may offer MD-driven insights into these connections, and it would be fruitful to leverage these tools to study Brownian motion in colloidal solutions and active matter.\cite{Chakraborty2011-yh, Jung2017-sb, Lee2019-ni, Seyler2019-dg, Seyler2020-kj, Goychuk2019-gl, Goychuk2020-ty, Diaz2021-ea, Cherayil2022-ns, Spiechowicz2022-oz}

We remark that the non-Gaussian behavior of the VACF at intermediate times, which switches from damped oscillations to pure exponential decay, coincides with the emergent exponential ranges in $\Fs$ and $\CT$. It is therefore possible that the hydrodynamic VACF formulation can lend dynamical insight into the liquid-vapor phase transition not otherwise available through other methods. Realizing a fully capable VACF theory would, however, require a means (i.e., a sum rule\cite{Resibois1977-mh, Henderson1992-zk}) to determine $a$ as a continuous function of density (and temperature) through the phase transition without relying solely on the short-time sum rule involving $\omega_E$ [req.~(5)], which is ill-defined at lower densities. Our methodology is nevertheless amenable to the use of different (kinetic) models used to determine model parameters (e.g., our use of Enskog theory for gaseous and supercritical argon calculations), which may be useful for probing fundamental questions pertaining to the behavior of supercritical\cite{Pedersen2008-jc, Ohtori2017-dj, Fomin2018-bb} and supercooled fluids.\cite{Pastore1988-tp, Barrat1990-kz, Balucani1990-rg, Balucani1990-zu, Puertas2007-ui, Baity-Jesi2019-rz, Ren2021-er, Levashov2013-oq, Ozawa2023-yl}

Extensions to the present work include deriving $q(ka)$ from first principles, as well as (numerically) solving the full zero-frequency constraint for $\DT$ [req.~(7)] at all densities and third-order ODE for $\CL(k,t)$ (i.e., beyond the relaxation limit). However, we expect the present theory to be valuable well beyond VACF calculations. For example, \citet{Alder1970-kg} and, more recently, \citet{Han2018-lc} and \citepunc{Lesnicki2017-bm}{,} have convincingly demonstrated that collective motions in discrete-particle fluids are well represented by hydrodynamic flow fields down to the single-particle scale. This suggests that stochastically driven R10 equations,\cite{Seyler2017-tm} which generalize the Landau-Lifschitz Navier-Stokes equations,\cite{Landau1966-ke} would represent a viable molecular-hydrodynamic model capable of reproducing molecular-scale flows.

Indeed, the efficacy of our VACF formulation rests largely on $q(ka)$. Even apparently subtle differences in the \emph{shape} of $q(ka)$ substantially alter the character of the VACF, e.g., the oscillatory nature for liquid rubidium or ``plateau'' for liquid argon (data not shown). The details of the VACF depend on the manner in which the fluid's distribution of kinetic energy transitions from (approximate) equipartition at long wavelengths to zero below the molecular scale. The kinetic energy distribution must, in turn, depend on the intermolecular potential, which also determines the radial distribution function, $g(r)$, or static structure factor, $S(k)$. However, whereas $g(r)$ or $S(k)$ characterizes static \emph{density} correlations (zeroth velocity-moment), $q(k)$ characterizes static \emph{momentum} correlations (first velocity-moment). Given its pivotal role in capturing molecular-scale behavior, it is thus our belief that the static spectrum is key to describing how continuum hydrodynamic modes emerge from the molecular scale and bridging the continuum and discrete-particle perspectives.

%


\begin{acknowledgments}
The authors are grateful for valuable discussions with Mark A. Hayes, Dmitry Matyushov, Jason Hamilton, Ralph V. Chamberlin, Paul Campitelli, and Kyle L. Seyler. SLS would like to warmly acknowledge Oliver Beckstein and the Blue Waters Graduate Fellowship program---a part of the Blue Waters sustained-petascale computing project supported by the National Science Foundation (awards OCI-0725070 and ACI-1238993) and the state of Illinois---whose generous support helped nucleate this research; Blue Waters is a joint effort of the University of Illinois at Urbana-Champaign and its National Center for Supercomputing Applications. CES was supported by the National Nuclear Security Administration Stewardship Sciences Academic Programs under Department of Energy Cooperative Agreement No. DE-NA0003764.
\end{acknowledgments}

\section*{Author Declarations}

\subsection*{Conflict of Interest}
The authors have no conflicts to disclose.

\subsection*{Author Contributions}
All authors contributed equally to this work.


\section*{Data Availability}
The data that support the findings of this study are available within the article and its supplementary material.

\renewcommand\thefigure{\thesection.\arabic{figure}}
\appendix

\section{Dynamic spectral densities}\label{sm:dynamic_spectral_density}
The dynamic structure factor $J(k,\omega)$ is the cosine transform of the current correlation function.

\begin{equation}
    J(k,\omega) = 2\int_0^\infty\!\! dt\,C(k,t)\cos(\omega t)
\end{equation}
Recalling that $\upsilon_k \equiv \upsilon + k^2 \mcd_\perp$ and $\gamma_k \equiv \gamma + k^2 D_\parallel$, we find for the transverse and longitudinal correlations respectively
\begin{equation}
    J_\perp(k,\omega) = \sdfrac{2k^2\vt^2\upsilon_k}{\bigl(\omega^2-k^2\vt^2\bigr)^2+\omega^2\upsilon_k^2}
\end{equation}
\begin{equation}
    J_\parallel(k,\omega) = \sdfrac{2k^2 \vL^2 \gamma_k}{\bigl(\omega^2-k^2\vL^2\bigr)^2+\omega^2k^4 \mcd_\parallel^2}
\end{equation}
Note the following property: as $\smash{\upsilon_k,\gamma_k \to 0}$ and $\smash{\mcd_\perp,D_\parallel \to 0}$, the spectral densities approach delta functions in the arguments $\omega=\pm k\vt$ and $\omega=\pm k\vL$.
Compare these results to purely viscous decay that one would obtain from the Navier-Stokes equation
\begin{equation}
    J_{NS}(k,\omega) = \sdfrac{2k^2\nu}{\omega^2+k^4\nu^2}
\end{equation}


\section{Determination of scale \texorpdfstring{\boldmath$a$}{a} for liquids [req.~(5)]}\label{sm:scale_a}

In sufficiently dense fluids such as a liquid, the Einstein frequency, $\omega_E$, has physical relevance and $a$ can be computed from the second-moment condition by considering the short-time expansion of the VACF
\begin{equation}\label{eq:sm_short_time_formal}
  \psi(t) = 1-\sdfrac{1}{2}\left(\sdfrac{t}{\tau_c} \right)^2 + \cdots = 1-\sdfrac{1}{2}\omega_E^2t^2 + \cdots
\end{equation}
where $\smash{\tau_c = \omega_E^{-1}}$ is the timescale characterizing the initial (de)correlation of the VACF and the Einstein frequency, $\omega_E$, is formally defined via
\begin{equation}
    \omega_E^2 \equiv \sdfrac{4\pi n_0}{3m}\!\intzinf[r]{r^2 g(r)\left(\sdfrac{d^2\phi}{dr^2}+\fr{2}{r}\sdfrac{d\phi}{dr}\right)}
\end{equation}
where $g(r)$ is the radial distribution function and $\phi(r)$ is the intermolecular potential. For large Sc---assumed to be the case for liquids---the short-time expansion of the modal correlation function dominates that from the self-intermediate scattering function. That is, when $\smash{\nu \gg D_s}$, we have $\smash{\vt^2/\upsilon \gg \vt^2/\gamma_s}$ and then $\smash{\gamma_s \gg \upsilon}$. Thus, in the liquid state, $\Fs(k,t) \approx 1$ over the timescale $\tau_c$ and the initial decay of the VACF (and the location of the first zero-crossing) is controlled by the modal current correlation functions, $\CT$ and $\CL$.

The integrand of the hydrodynamic VACF formula, \cref{eq:hd_VACF}, can thus be approximated as
\begin{equation}
  \Fs(k,t)C(k,t) \approx 1-\sdfrac{1}{2}k^2\left(\sdfrac{1}{3}\vL^2+\sdfrac{2}{3}\vt^2\right)t^2
\end{equation}
and so the short-time VACF is approximately
\begin{equation}\label{eq:sm_short_time}
  \psi(t) \approx \sdfrac{\Np a^3}{3}\! \intzinf[k] k^2 q(ka) \left[ 1-\sdfrac{1}{2}k^2\left(\vL^2 + 2\vt^2\right)t^2 \right]
\end{equation}
Equating the quadratic terms in \cref{eq:sm_short_time_formal} and \cref{eq:sm_short_time}, we obtain
\begin{equation}\label{eq:sm_second_moment}
  \omega_E^2 = \sdfrac{\Np a^3}{3}\! \intzinf[k] k^4 q(ka) \left(\vL^2 + 2\vt^2\right)
\end{equation}
which is the second frequency-moment condition (i.e., $f$-sum rule).\cite{Hansen2013-rh}

To obtain \cref{eq:a_rescale}, explicitly evaluate \cref{eq:sm_second_moment} for the generalized Gaussian spectrum
\[
  \psi(t) = \sdfrac{p\!\:a^3}{8^{1/p}\Gamma(3/p)}\!\intzinf[k] k^2\,e^{-\frac{1}{2}|ka|^p}\Fs(k,t)C(k,t)
\]
The scale $a$ can now be deduced from the correlation time $\tau_c$ through the second wavenumber moment of the static spectrum, $q(k)$:
\begin{equation}\label{seq:einstein_freq}
  \omega_E^2 = \sdfrac{p\,a^3\rv_a^2}{8^{1/p}\Gamma(3/p)}\!\intzinf[k] k^4\,e^{-\frac{1}{2}|ka|^p}
\end{equation}
where $\rv_a^2 \equiv \bigl(\vL^2+2\vt^2\bigr)/3$. Carrying out the integral analytically and rearranging to solve for $a$ yields
\begin{equation}\label{seq:a_rescale}
   a = \sqrt{\sdfrac{4\smash{^{1/p}}\Gamma(5/p)}{\Gamma(3/p)}} \sdfrac{\rv_a}{\omega_E}
\end{equation}
which agrees with \cref{eq:a_rescale}. For the specific case of a pure Gaussian static spectrum, $\smash{a = \sqrt{3}\rv_a/\omega_E}$. In the liquid state, it is seen that the main decay timescale is controlled by the molecular scale $a$ via $\smash{\tau_c = \omega_E^{-1} \sim a/\rv_a}$.

\section{Zero-frequency constraint for liquids [req.~(7)]}\label{sm:green_kubo_constraint}
\setcounter{figure}{0}
The Green-Kubo relation for self-diffusion is
\begin{equation}\label{eq:sm_green_kubo}
    D_s = \vt^2\!\intzinf[t] \psi(t)
\end{equation}
%
In the liquid regime, it is reasonable to take $\smash{\CL(k,t) \to 0}$ and $\smash{\Fs(k,t) \to 1}$, which allows us to evaluate \cref{eq:sm_green_kubo} analytically using \cref{eq:hd_VACF} to obtain
\begin{equation}
    D_s = \sdfrac{2}{3}\left[\DT + \chi(p)a^2\upsilon\right]
\end{equation}
where $\chi(p)$ is a coefficient that depends on the sharpness of the spectral cutoff. After rearranging, we obtain
\begin{equation}\label{eq:sm_analytic_DT}
    \DT = \sdfrac{3}{2}\Ds - \chi(p)a^2\upsilon
\end{equation}
which, as shown below, yields analytic constraints for $\DT$ when $D_s$ is known (e.g., experimentally) and the static spectrum is represented by either the generalized Gaussian or Lorentzian forms [\cref{eq:gauss_spectrum,eq:lorentz_spectrum}].

For the transverse current, \cref{eq:sm_green_kubo} becomes
\begin{equation}\label{eq:sm_transverse_green-kubo}
    D_s = \sdfrac{2}{3}\vt^2\mathcal{N}\! \intzinf[k] k^2 q(ka) \!\intzinf[t] \CT(k,t)
\end{equation}
Carrying out the time integration first yields
\begin{equation}
    \vt^2\!\intzinf[t] C_\perp(k,t) = \DT + \sdfrac{\upsilon}{k^2}
\end{equation}
so that the $k$-integral that remains to be evaluated is
\begin{equation}\label{eq:sm_qka_transverse}
   I_\perp\bigl[q(ka)\bigr] = \sdfrac{2}{3}\Np a^3\!\intzinf[k] k^2 q(ka)\left( \DT + \sdfrac{\upsilon}{k^2} \right)
\end{equation}
For a generalized Gaussian spectrum, \cref{eq:sm_qka_transverse} becomes
\begin{equation}\label{eq:sm_qGG_transverse}
    \sdfrac{3}{2} I_\perp\Big[e^{-\frac{1}{2}(ka)^p} \Big] = \DT + \sdfrac{1}{4^{1/p}}\sdfrac{\Gamma\big(1/p\big)}{\Gamma\big(3/p\big)}a^2\upsilon
\end{equation}
and, for a generalized Lorentzian,
\begin{equation}\label{eq:sm_qGL_transverse}
    \sdfrac{3}{2} I_\perp\Bigg[\sdfrac{1}{1+(ka)^p} \Bigg] = \DT + \sdfrac{\csc(\pi/p)}{\csc\big(3\pi/p\big)}a^2\upsilon
\end{equation}
Inspection of \cref{eq:sm_qGG_transverse,eq:sm_qGL_transverse} reveals that the transverse current correlation uniquely contributes a $p$-dependent coefficient of the $a^2\upsilon$ term, $\chi(p)$, unlike the longitudinal current correlation, as well as depending on the functional form of $q(ka) = q_p(ka)$:
\begin{equation}\label{eq:sm_qGL_transverse2}
    \sdfrac{3}{2}\Ds = \sdfrac{3}{2}I_\perp\bigl[q_p(ka)\bigr] = \DT + \chi(p) a^2\upsilon
\end{equation}
Thus, $\DT$ can be readily computed using \cref{eq:sm_qGG_transverse} or \cref{eq:sm_qGL_transverse} with \cref{eq:sm_analytic_DT}.

\begin{figure}[t]
  \centering
  \includegraphics[width=\columnwidth]{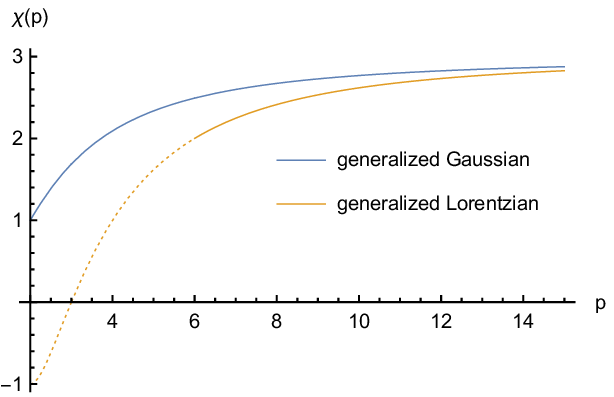}
  \caption{Coefficient $\chi(p)$ in \cref{eq:sm_qGL_transverse2} from the transverse component's contribution to the long-time diffusivity for generalized Gaussian, \cref{eq:gauss_spectrum}, and Lorentzian, \cref{eq:lorentz_spectrum}, static spectrum functions. The dashed line indicates values of $p$ for which the second-moment $k$-integral [req.~(5a)] is divergent for the generalized Lorentzian spectrum (i.e., $p < 6$).}
  \label{fig:sm_transverse_coeff_of_static_spectra_vs_p}
\end{figure}

\Cref{fig:sm_transverse_coeff_of_static_spectra_vs_p} compares the magnitude of $\chi(p)$ for the generalized Gaussian (blue) and generalized Lorentzian (orange) forms for the static spectrum.
It is seen, for instance, that a generalized Gaussian with $p = 4$ gives roughly the same contribution to the overall diffusivity as the generalized Lorentzian with $p \approx 6$.

\section{Parameter determination for dilute fluids}\label{sm:gas_parameters}

\paragraph*{\bf Regularization and the exponential range [req.~(4)].}
At lower densities, self-diffusion becomes important and $\Sc$ is relatively small as compared to the liquid case. For a dilute gas, where $\smash{\Sc \sim \bigO(1)}$, it would be reasonable to assume the exponential range [req.~(4)] arises only from the product $\Fs(k,t)\CT(k,t)$. Then, one might define an \emph{effective} decay rate, $\Omega_0$, to be the sum of the SISF and transverse current collision rates, i.e., $\Omega_0 = \gamma_s + \upsilon$. It will be shown below that this assumption is not unreasonable and can be formally justified by enforcing req.~(7) using large-$k$ approximations for the correlation functions. Specifically, the exponential range should arise for values of $k$ where $k \vt > \tau$ and $\Omega_0 = \tau^{-1}$ [req.~(4)]. Indeed, unlike in the liquid case, the SISF cannot be treated as approximately constant (i.e., $\smash{\Fs \neq 1}$) over the time interval with the dominant contribution to self-diffusion, which precludes an exact analytic integration over $k$. However, for dilute monatomic gases, the dominant contribution to $\Ds$ generally comes from the large-$k$ forms of $\CT(k,t)$ and $\Fs(k,t)$, in the Green-Kubo relation allowing an approximate $k$ integration.

In this scenario, regularization plays a critical role, as seen by considering the slowest relaxation rate of $\CT(k,t)$---the smaller of the two exponents in \cref{eq:CTk}, which, for large $k$, is
\begin{equation}\label{eq:exp_decay}
   -\sdfrac{1}{2}\upsilon_k + \sdfrac{1}{2}\sqrt{\smash[b]{\upsilon_k^2-4k^2\vt^2}} \sim -\sdfrac{k^2\vt^2}{\upsilon+k^2\DT} \sim -\sdfrac{\vt^2}{\DT}
\end{equation}
This relaxation rate must be finite as $k\to\infty$ [req.~(4)], which, evidently, necessitates the inclusion of the regularization diffusion coefficient, $\DT$. A similar argument holds for the exponential range of $\Fs(k,t)$, where $\Ds$ is the corresponding regularization coefficient.

Similarly, the longitudinal relaxation rate for large-$k$ is $\smash{{\sim}\vL^2/\mu}$; however, the bulk viscosity of a dilute monatomic gas is typically very small compared to its shear viscosity,\cite{Meier2005-ap, Jaeger2018-pg, Sharma2023-qt} making its contribution to the diffusivity small as well. To see this analytically, consider the VACF with a Gaussian static spectrum ($\smash{p=2}$) and bulk viscosity set to zero: the longitudinal contribution is the rapidly decaying form $\psi_\parallel(t) \sim \smash{(1-\vL^2t^2/a^2)\exp\bigl(-\vL^2 t^2/2a^2\bigr)}$, which integrates exactly to zero. Thus, while $\CL$ can affect the short-time VACF \emph{structure}, only $\CT$ and $\Fs$ dictate the exponential range and overall diffusivity under dilute conditions.

\paragraph*{\bf Zero-frequency constraint [req.~(7)].}
Excluding the longitudinal component, the Green-Kubo relation for large $k$ becomes [req.~(7)]
\begin{equation}\label{eq:sm_gas_green-kubo}
    D_s = \sdfrac{2}{3}\vt^2\left(\gamma_s+\sdfrac{\vt^2}{\DT}\right)^{-1}
\end{equation}
Note the appearance of the additional parameter $\gamma_s$ due to the treatment of finite ${\rm Sc}$. To uniquely determine the parameters, we now require a separate relation between $\DT$ and $\upsilon$ or another way to determine $\gamma_s$. At present, it seems reasonable to assume $\smash{\DT \to \vt^2/\upsilon}$, which is consistent with the assumption that $\smash{D_s \to \vt^2/\gamma_s}$, which was used in \crefrange{eq:self_density}{eq:SISF_equation}---the equations for the self-density and SISF. We remark that when $\smash{\DT = \vt^2/\upsilon}$, the roots of \cref{eq:CT_ode} factor nicely to give
\begin{equation}\label{eq:sm_factored_CT}
    \CT(k,t) = \fr{\upsilon e^{-k^2\DT t}-k^2\DT e^{-\upsilon t}}{\upsilon-k^2\DT }
\end{equation}
matching the neat form of \cref{eq:SISF_equation} and preserving the positivity of the VACF in accordance with what is expected for a dilute gas.

Setting $\smash{\DT = \vt^2/\upsilon}$ in \cref{eq:sm_gas_green-kubo}, we now have
\begin{equation}\label{eq:sm_gas_green-kubo_2}
    D_s = \sdfrac{2}{3}\sdfrac{\vt^2}{\gamma_s+\upsilon}
\end{equation}
Clearly, we cannot directly set $\smash{D_s=\vt^2/\gamma_s}$ on the left-hand side of \cref{eq:sm_gas_green-kubo_2}, as there would be no positive solution for $\upsilon$. We instead assume, as in the liquid case, that $\gamma_s$ represents the \emph{rescaled} (self-)collision frequency, while $\gamma_{s0}$ is its corresponding \emph{base} value determined by $\smash{D_s \equiv \vt^2/\gamma_{s0}}$. With this assumption, \cref{eq:sm_gas_green-kubo_2} becomes
\begin{equation}\label{eq:sm_gas_green-kubo_3}
    \gamma_{s0}=\sdfrac{3}{2}\left(\gamma_s+\upsilon\right)
\end{equation}

\paragraph*{\bf Parameter rescaling [req.~(6)].}
Finally, we require that the Schmidt number be invariant after rescaling: $\smash{\Sc \equiv \nu/\Ds = \gamma_{s0}/\upsilon_0=\gamma_s/\upsilon}$, which fixes the ratio of the rescaled collision frequencies to the ratio of their base values. Substituting $\upsilon = \gamma_s/\Sc$ into \cref{eq:sm_gas_green-kubo_3} and solving for $\gamma_s$ leads to the condition
\begin{equation}\label{eq:sm_gas_gamma_s}
  \gamma_s = \sdfrac{2}{3}\sdfrac{\Sc}{\Sc+1} \gamma_{s0} = \left(\sdfrac{a_0}{a}\right)^2 \gamma_{s0}
\end{equation}
where the second equality is obtained by recalling that $a$ is determined by preserving the long-time asymptotic form of the VACF [req.~(6)], which implies
\begin{equation}\label{eq:sm_gas_a}
    a^2 = \sdfrac{3}{2}\Bigl(1+\Sc^{-1}\Bigr)a_0^2
\end{equation}
Thus, $\gamma_s$, $\upsilon$, and $\gamma$ may be determined, respectively, via $\smash{\gamma_{s0} = \vt^2/\Ds}$, $\smash{\upsilon_0 = \vt^2/\nu}$, and $\smash{\gamma_0 = \vt^2/\mu}$ given known inputs for the transport coefficients ($\Ds$, $\nu$, and $\mu$).

\paragraph*{\bf Parameters for gaseous and supercritical argon.}
Given the sparsity of numerical and experimental data for transport coefficients of dilute monatomic gases, the VACFs in \cref{fig:gaseous_argon} were produced using transport coefficients calculated from Enskog theory,\cite{Chapman1990-rl, Erpenbeck1991-te, Heyes2022-gf} supplemented by isothermal data for argon (at \SI{130}{\kelvin} and \SI{300}{\kelvin}) from the NIST Chemistry WebBook. For the transport coefficients, we took $\smash{D_s \to D_E = \vt^2/\gamma_{s0}}$, $\smash{\nu\to\nu_E = \vt^2/\upsilon_0}$, and $\smash{\mu\to\nu_{bE} = \vt^2/\gamma_0}$, where
\begin{align}
  D_E &\equiv 1.01896\sdfrac{D_0}{g(\sigma)}\label{eq:sm_enskog_D}\\
  \nu_E &\equiv 1.016\,\nu_0 \left[\sdfrac{1}{g(\sigma)} + 0.8(b n_0) + 0.7615(b n_0)^2g(\sigma) \right]\label{eq:sm_enskog_eta}\\
  \nu_{bE} &\equiv \sdfrac{16}{5\pi}\nu_0 (bn_0)^2g(\sigma)\label{eq:sm_enskog_mu}
\end{align}
are the Enskog diffusion coefficient, (kinematic) shear viscosity, and (kinematic) bulk viscosity, respectively.\cite{Heyes2022-gf}

In \crefrange{eq:sm_enskog_D}{eq:sm_enskog_mu}, $n_0$ is the equilibrium number density, $\sigma$ the hard-sphere diameter, and $b \equiv (2/3)\pi \sigma^3$ is the second virial coefficient for a hard-sphere fluid. $D_0$ and $\nu_0$ are the values of the self-diffusion and shear viscosity coefficients in the limit of zero density,
\begin{align}
  D_0 &\equiv \sdfrac{3\vt}{8\!\sqrt{\pi}n_0\sigma^2}\\
  \nu_0 &\equiv \sdfrac{5\vt}{16\!\sqrt{\pi}n_0\sigma^2}
\end{align}
and $g(\sigma)$ is the radial distribution function evaluated at the hard-sphere point of contact. Analytic approximations for $g(\sigma)$ can be obtained from the Percus-Yevick equation, scaled particle theory,\cite{Reiss1959-mi, Reiss1960-lp, Helfand1960-mi} or the Carnahan-Starling equation of state;\cite{Carnahan1969-al, Song1989-qp, Sigurgeirsson2003-dw} we used the Carnahan-Starling approximation for the 3D pair distribution:
\begin{equation}
    g(\sigma) \approx \sdfrac{1-\phi/2}{\bigl(1-\phi\bigr)^3}
\end{equation}
where the packing fraction $\smash{\phi \equiv (\pi/6)n_0 \sigma^3}$.

For all calculations in \cref{fig:gaseous_argon}, we used a Gaussian static spectrum ($\smash{p=2}$) primarily as proof-of-principle, though one should expect higher values of $p$ to apply only at the highest densities; e.g., in the liquid state, where propagating longitudinal and shear wave modes may be significant for softer intermolecular potentials. We also took $\smash{\sigma = \SI{3.405}{\angstrom}}$ (argon Lennard-Jones diameter), $m = \SI{40}{\dalton}$ for the (atomic) mass, and set $\DT = \vt^2/\upsilon$ (as discussed above) and $\DL = 4\mu/3$ (as in the liquid case). We determined $\vL^2$ from the adiabatic index, i.e., $\smash{\vL^2/\vt^2} = C_P/C_V$, using heat capacity values obtained from the NIST Chemistry WebBook for each combination of temperature and density, $T$ and $\rho_0$, where $\rho_0 = mn_0$.

\section{Comparison with MD calculations for a supercritical Lennard-Jones fluid}\label{sm:sc_argon}
\setcounter{figure}{0}
In \citenst{Lesnicki2016-bc}, MD simulations are performed for a supercritical Lennard-Jones (LJ) fluid at a reduced density $\smash{n_0^* = n_0 \sigma^3 = 0.5}$ and reduced temperature $\smash{T^* = \kt/\epsilon = 1.5}$, where $\sigma$ and $\epsilon$ are the LJ diameter and energy, respectively. To compare with the results of \citenst{Lesnicki2016-bc}, we took $\smash{m = \SI{40}{\dalton}}$, $\smash{\sigma = \SI{3.405}{\angstrom}}$, and $\smash{\epsilon = \SI{120}{\kelvin}}$, so that the (dimensional) mass density and temperature are, respectively, $\smash{\rho = m n_0 = \SI[per-mode=symbol]{841.3}{\kilogram\per\cubic\meter}}$ and $\smash{T = \SI{180}{\kelvin}}$; note that, for this temperature-density combination, $\smash{P \approx \SI{19.2}{\mega\pascal}}$. For $\rv_\parallel$, we again use the adiabatic index computed from heat capacities obtained from the NIST Chemistry WebBook, SRD 69 for isothermal properties of argon (i.e, $\smash{\vL^2/\vt^2 = C_P/C_V}$). To obtain base values for the collision frequencies, i.e., $\upsilon_0$, $\gamma_0$, etc., we used the same Enskog diffusion and viscosity coefficient formulas given in \cref{sm:gas_parameters}. However, we chose to treat this comparatively dense supercritical (SC) state slightly differently, as the packing fraction is $\phi \approx 0.262$---more than half of argon's triple point (TP) density ($\phi \approx 0.445$). In particular, we employed a hybrid treatment (detailed below) wherein the molecular scale $a$ was computed via the second-moment condition [req.~(5) and \cref{seq:a_rescale}] using simple physical arguments to obtain an estimated Einstein frequency of $\omega_E = \SI{4.75}{\per\pico\second}$. The resulting VACF---\cref{fig:sm_vacf_dense_sc_argon}, blue---is in striking agreement with the MD calculations shown in Fig.~1 of \citenst{Lesnicki2016-bc}. Our estimate for the Einstein frequency is based on the following calculations and physical reasoning.

\begin{figure}[t]
  \centering
  \includegraphics[width=\columnwidth]{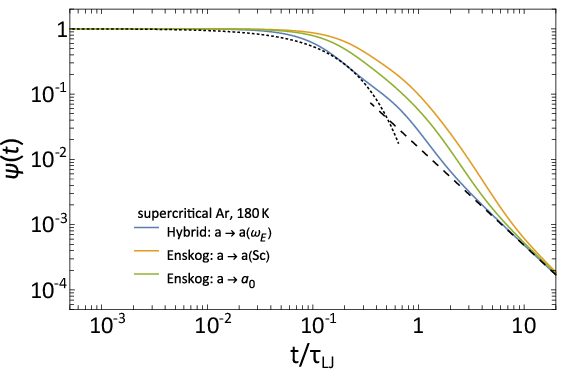}
  \caption{VACFs corresponding to a dense supercritical Lennard-Jones fluid studied in \citenst{Lesnicki2016-bc} (cf. Fig.~1). The VACF from the hybrid procedure described in this appendix (blue) is depicted with VACFs produced for two different (larger) values of $a$: setting $\smash{a \to \sqrt{(3/2)(1+\Sc^{-1})}\!\:a_0}$ via \cref{eq:sm_gas_a} (orange) and explicitly fixing $a \to a_0$ (green).}
  \label{fig:sm_vacf_dense_sc_argon}
\end{figure}

First, at $\smash{\phi \approx 0.262}$, the mean interparticle separation (center-to-center) is $\smash{\dip{SC}_0 \sim n_0^{-1/3} \approx \SI{4.29}{\angstrom}}$, so the contact (surface-to-surface) distance, $\dip{SC}$, may be estimated as $\smash{\dip{SC} \approx \dip{SC}_0 - \sigma \approx \SI{0.885}{\angstrom}}$; likewise, $\smash{\dip{TP} \approx \SI{0.190}{\angstrom}}$ at the TP. It is within reason to assume the Einstein frequency, $\omega_E$, is physically relevant for SC argon at the given density because the surface-to-surface spacing is (still) substantially smaller than the particle diameter; i.e., we assume that particles cannot easily ``break through the cage'' formed by its neighbors at this (packing) density. Second, to arrive at an estimate for $\omega_E$, we assume that the larger interparticle spacing in the SC case leads to an increased oscillation period, $\Delta\tau$, over that of TP argon, where $\smash{\Delta\tau = \taup{SC} - \taup{TP}}$, and $\smash{\taup{SC} \equiv 2\pi/\omega_E^\text{\tiny SC}}$ and $\smash{\taup{TP} \equiv 2\pi/\omega_E^\text{\tiny TP}}$ are the respective oscillation periods. Finally, to estimate the increase in oscillation period, we make the simple physical assumption that $\Delta\tau$ is an intervening \emph{ballistic} interval arising from the increased surface-to-surface distance, an additional separation of $\smash{\dip{SC} - \dip{TP} \approx \SI{0.695}{\angstrom}}$ over the TP case. This leads to $\smash{\Delta\tau \approx 2\bigl(\dip{SC} - \dip{TP} \bigr)/\rv_b}$, where the factor of 2 accounts for two ballistic traversals per cage oscillation and we take the ballistic speed to $\smash{\rv_b \approx \sqrt{\smash[b]{2\kt/m}} = \sqrt{\smash[b]{2}}\rv_0 }$ (i.e., the most probable speed of a Maxwell-Boltzmann distribution).

Putting everything together, we arrive at a rough estimate for the oscillation period for the dense SC fluid:
\begin{align*}
    \taup{SC} = \sdfrac{2\pi}{\omega_E^\text{\tiny TP}} + \sdfrac{2\bigl(\dip{SC} - \dip{TP} \bigr)}{\sqrt{\smash[b]{2\rv_0^2}}}
        \approx \sdfrac{2\pi}{\SI{7.7}{\per\pico\second}} + \sdfrac{2\SI{0.695}{\angstrom}}{\SI[per-mode=symbol]{2.74}{\angstrom\per\pico\second}} \approx \SI{1.32}{\pico\second}
\end{align*}
which yields for the Einstein frequency
\begin{equation}\label{eq:sm_einstein_freq_dense_sc}
    \omega_E^\text{\tiny SC} = \sdfrac{2\pi}{\taup{SC}} \approx \SI{4.75}{\per\pico\second}
\end{equation}
Finally, we use a generalized Gaussian spectrum with $\smash{p \approx 1.5}$ for which \cref{seq:a_rescale} gives $\smash{a \approx \SI{2.28}{\angstrom}}$ and yields excellent agreement (\cref{fig:sm_vacf_dense_sc_argon}, blue) throughout the timescales sampled by MD. Note that time is expressed in LJ reduced units, $\smash{t^* \equiv t/\tau_\text{LJ}}$, where $\smash{\tau_\text{LJ} \equiv \sqrt{m \sigma^2/\epsilon}}$ is the characteristic LJ timescale; for argon, $\smash{\tau_\text{LJ} \approx \SI{2.156}{\pico\second}}$. \Cref{fig:sm_vacf_dense_sc_argon} also shows VACFs computed using two different (larger) values of $a$: computing $a$ from \cref{eq:sm_gas_a} for dilute fluids (orange) and explicitly setting $\smash{a \to a_0}$ (green). Note that the blue curve reproduces the rather subtle ``plateau'' region ($\smash{t^* \approx 0.15\text{--}0.6}$) due to the sound-like mode originating from the longitudinal component, as well as the nuanced transition from exponential-like to diffusive decay for $\smash{t^* \gtrsim 1}$.






\Urlmuskip=0mu plus 1mu\relax
\bibliographystyle{apsrev4-2}
\bibliography{inputs/references}

\end{document}